\documentclass[a4paper]{article}

\usepackage[english]{babel}
\usepackage[utf8x]{inputenc}
\usepackage[T1]{fontenc}

\usepackage{epsfig}

\usepackage{amsmath}
\usepackage{graphicx}
\usepackage[colorinlistoftodos]{todonotes}

\usepackage{soul} 
\usepackage{comment} 

\begin{document}

\title{Improving User's Experience \\through Simultaneous Multi-WLAN Connections}
\author{Guillem Cañizares, Boris Bellalta\\Wireless Networking Group \\ Universitat Pompeu Fabra, Barcelona\\guillem.canizares@upf.edu, boris.bellalta@upf.edu}

\maketitle

\begin{abstract}
In highly-dense IEEE 802.11 deployments, areas covered by multiple Wireless Local Area Networks (WLANs) will be common. This opens the door for stations equipped with multiple IEEE 802.11 interfaces to use several WLANs simultaneously, which not only may improve user experience, achieving a better connection with higher throughput and resilience; but it may also improve the network utilization. In this paper we investigate such a scenario. First, using a test-bed, consisting of a single station equipped with two interfaces and two access points, we observe that the file transfer time between the station and a destination server can be significantly reduced, studying with special attention the case in which both links do not have the same available bandwidth. Then, using a Markovian model that captures the scenario's dynamics in presence of multiple stations, we observe that in addition to improve individual station's performance, we can also improve the utilization of a multi-Access Points network despite increasing the contention level.
\end{abstract}
 




\section{Introduction}

Future Wireless Local Area Networks (WLAN) scenarios presume the existence of multiple overlapping IEEE 802.11 WLANs over the same area \cite{bellalta2016ieee,bellalta2016next}. In those places under coverage from different Access Points (APs), the use of multiple IEEE 802.11 interfaces --physical or virtual-- at the same time can be a promising solution to improve both the user experience and network utilization. 

To the best of our knowledge, we have not found any other paper focusing on a similar scenario. However, there is a large number of works that consider the existence of multiple wireless interfaces in the stations. For example, Brik et al. propose a multiple radio system in \cite{brik2005eliminating} to eliminate the handoff latency in WLANs, reducing it to around 30-40 ms using their \textit{MultiScan} approach. 

There are also radio virtualization proposals such as the Picasso \cite{hong2012picasso} project. Authors propose a radio design that allows simultaneous transmission and reception on separate spectrum fragments, using a single RF front end and antenna. They provide evidences that their prototype can virtualize a single radio into separate independent frequency slices, and achieve the same cumulative throughput than using the same number of real interfaces. Furthermore, authors in \cite{al2011virtualization} stand that the use of virtual wireless interfaces to connect to multiple networks saves energy, minimizes the physical space, and improves the coexistence in dense deployments.



Based on this, the incorporation of a transport protocol that is able to use different interfaces simultaneously, such as Multipath TCP (MPTCP), would become crucial in order to provide a better user experience. Arzani et al. \cite{arzani2014impact} expose that the use of a congestion controller and packet schedulers are basic for MPTCP performance, while path election and buffer sizes have also a significant impact. The authors of \cite{paasch2014experimental} analyze the benefits that MPTCP can provide to a wireless connection. They test single path, 2 and 4 Multi Path connections and observe the impact of the flow size on the average latency, concluding that the latency achieved by MPTCP is comparable to the smallest latency produced by either Wi-Fi or LTE connections in single path, except for small files under MB size. 

In this paper, instead on focusing on the performance of multiple paths to simply transmit data, we focus on a more fundamental problem. We investigate if the use of multiple IEEE 802.11 interfaces, and simultaneous connections to multiple APs, can be a suitable solution to improve the overall network performance in dense scenarios, providing some evidences about its potential benefits and drawbacks. Namely, 
\begin{enumerate}
	\item For the case of a single station, we study through experiments the performance gains when multiple IEEE 802.11 interfaces placed in a single node are used simultaneously, considering unbalanced links and background traffic.
    \item We develop a simple analytical model to quantify the performance gains when multiple interfaces are used in scenarios with multiple stations. 
\end{enumerate}

The rest of the paper is organized as follows. Section \ref{Sec:Experiments} introduces the Testbed, the Java Application developed and the experiments done. The tests performed and the obtained results are extensively explained. In section \ref{Sec:Analysis} we show the analytical results for scenarios with multiple stations. Finally, we present the conclusions in section \ref{Sec:Conclusions}.


\section{Test-bed and Experiments} \label{Sec:Experiments}

\subsection{Test-bed}

\begin{figure}[h]
\centering
\epsfig{file=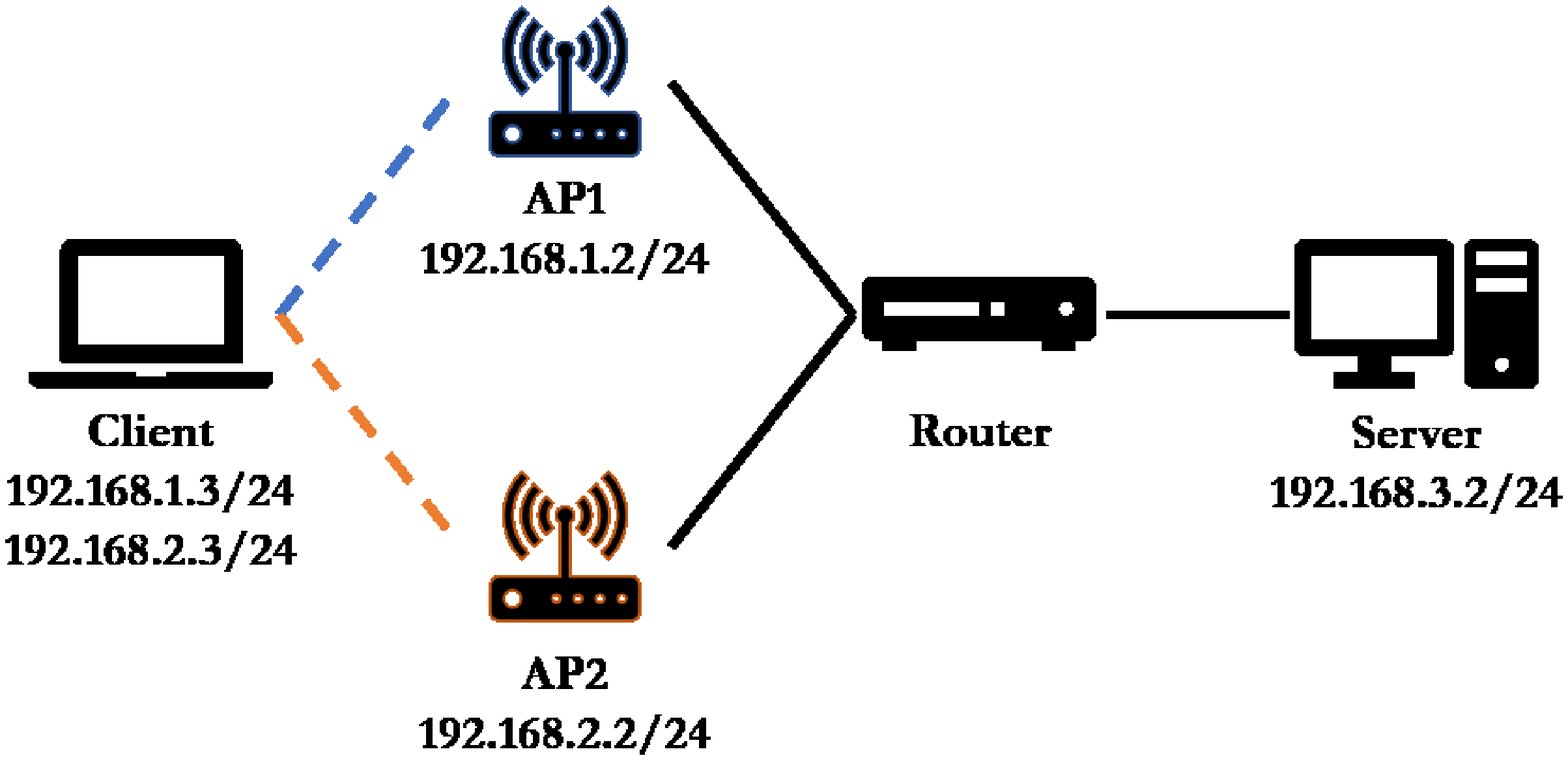,width=0.99\columnwidth,angle=-0}
\caption{Testbed}
\label{Fig:Setup}
\end{figure}

Figure \ref{Fig:Setup} shows the testbed we used for the experiments. It consists of a server (HP Compaq 600 pro) and a client (DELL Latitude 5580) connected through a router and two wireless access points -AP1 and AP2- (TP-Link AC1750). The client uses two wireless interfaces (Intel Dual Band Wireless-AC 8265 and Realtek TP-Link TL-WN822N). Each wireless access point is located 1 meter away of the client, being separated the same distance between them. The wireless system operates using the 802.11n standard. The Server and both APs are directly connected to the router. Each one of them belong to different subnets.

Ubuntu 16.04.3 LTE (Xenial Xerus) is installed on both client and server. The two APs use channels 1 and 11 respectively, minimizing any interference between them. The average signal strength measured at the client’s device is $-33$ dBm in both interfaces. The RTT between the client and the server is less than $5$ ms for both of them. Our testbed is not connected to the Internet to avoid the presence of external traffic that could have an impact on the results. 

\begin{figure}[h]
\centering
\epsfig{file=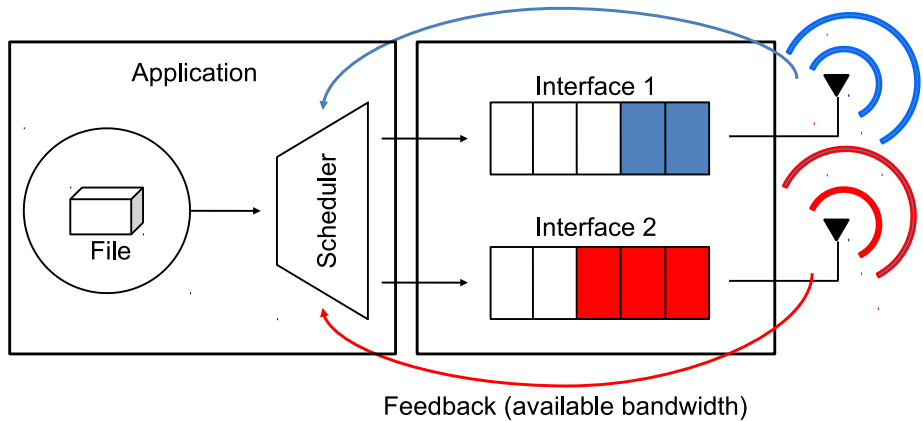,width=0.99\columnwidth,angle=-0}
\caption{Illustration of the client operation}
\label{Fig:Multi-Interfaces}
\end{figure}

Figure \ref{Fig:Multi-Interfaces} illustrates the client operation. The application, besides generating data, distributes the packets to the corresponding interfaces through a scheduler. The scheduler splits the file in different chunks and distributes them between all the interfaces based on the instantaneous available bandwith of each link. For this to happen, information about of the state of the network is sent to the application.

\subsection{Application}

We have developed a Java Application (APP) which connects the client and the server through a multi-socket connection. An independent thread is used to control the data transmission for each active wireless interface. Interfaces are bound to the correspondent IP address and server port number. The APP sends files from the client to the server, and returns the file transfer time. The files can be sent using one or two interfaces following different splitting ratios, thus generating multiple data chunks, distributing them to both interfaces. Using a $n$ split ratio means that the file is divided in $n$ chunks. Then, the first chunk is sent through the slow link, and the remaining $n-1$ are sent through the fast link.


\subsection{Experiments}

We focus on the file transfer time when files of different sizes are sent from the client to the server. We designed a test bench that included a total of ten files which size increases from 1 MB to 128 MB. Every test --i.e., a single file upload-- is reproduced ten times in order to obtain reliable results. Three different experiments were done:

\begin{enumerate}
\item \textbf{Upload Test}: test files are split 50/50 (i.e., $n=2$) and uploaded to the server. The total file transfer time using two interfaces is evaluated and compared to the case in which a single interface is used.
\item \textbf{Different Link Throughputs Test}: test files are uploaded to the server when the achievable throughput in one of the paths is significantly lower than in the other. The total file transfer time using a 50/50 splitting ratio is compared to the case in which the best splitting ratio is used.
\item \textbf{Background Traffic Test}: test files are uploaded to the server in different network conditions, as one of the links carries background traffic sent from the client to the server. The total file transfer time using a 50/50 splitting ratio is compared to the case in which the best splitting ratio is used.
\end{enumerate}

\subsection{Results}

In this section we present and discuss the obtained results in the tests presented above.

\subsubsection{Upload Test}

\begin{figure}[t!!!!!!!!]
\centering
\epsfig{file=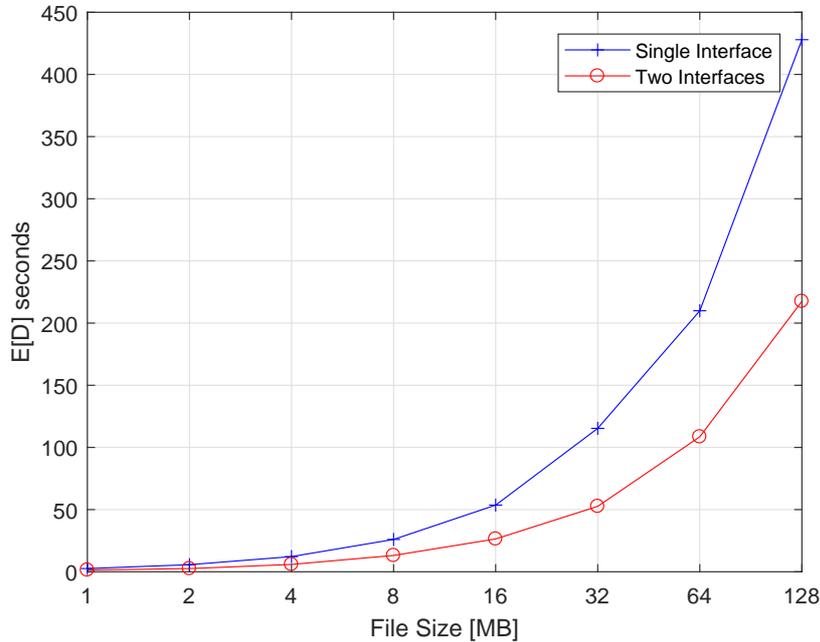,width=1\columnwidth,angle=-0}
\caption{File Transfer Time (First Test)}
\label{Fig:Test1}
\end{figure}

In this first test, files are first uploaded to the server using only a single link (the slower one, Link 2). Then, a 50/50 split is performed to the file and each file chunk is uploaded to the server through a different interface. Figure \ref{Fig:Test1} shows the file transfer time when a single and two interfaces are used. For instance, for the 128 MB size, we observe that the use of 2 interfaces reduces the transfer time more than 3 minutes. 

We observe that the ratio between the file transfer time when a single and two interfaces are used is always around 0.5. This means that the file transfer time using two interfaces lasts closely half of the time using a single interface. 

\subsubsection{Different Link Throughputs Test}

In this second test, files are uploaded to the server in different network conditions. Link 1 throughput calculations estimate an average of 2 Mbps, while Link 2 works 6 times faster, around 12 Mbps. Calculations are computed using \textit{iperf}\footnote{https://iperf.fr/} and \textit{ifstat}\footnote{http://gael.roualland.free.fr/ifstat/} tools. Files are first split 50/50 to observe how the presence of different transmission rates affect the file transfer time. Then, a second split ratio is applied in order to achieve a similar time in both links. Two different file sizes are tested in this scenario: 16 MB and 64 MB.

\begin{figure}[t!!!!!!!!]
\centering
\epsfig{file=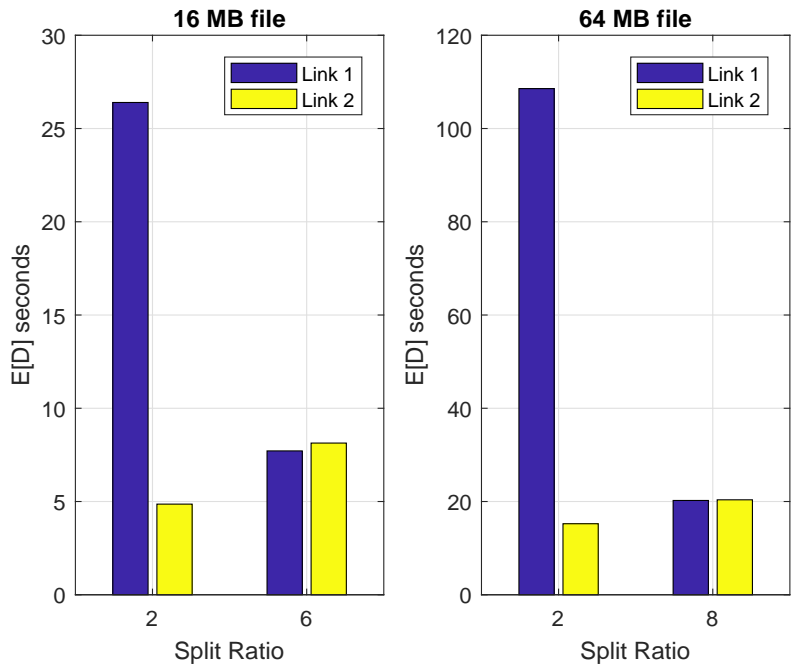,width=1\columnwidth,angle=-0}
\caption{File Transfer Time for different file sizes (Second Test)}
\label{Fig:Test2}
\end{figure}

Figure \ref{Fig:Test2} shows the results for the second test. The client transmits the 16 MB file using both links, sending through each one 8 MB --i.e., half of the file-- in a 50/50 split. It can be seen that the chunk transmitted through the slower link has a transfer time far larger than the time of the chunk transmitted through the fast link. To reduce this difference, a second splitting ratio is applied. This time, we split the file using a splitting factor of $n=6$, sending the 83.3\% of the file through the fast link, while the remaining 16.7\% is sent through the slow link. We observe that using this fairer split we achieve a 71\% reduction in the file transfer time. However, Link 2 transfer time is incremented by 67\%, even though this increment does not harm the total connection time. Similar results are obtained for the 64 MB file. As in the previous case, using the best splitting ratio --i.e., $n=8$-- reduces the file transfer time in a 81\% and increments Link 2 by 34\% respectively.

\subsubsection{Background Traffic Test}


In this third test, TCP throughput values for both links are between 10 and 12 Mbps. However, one of the two links also carries background traffic. The background traffic is generated using the \textit{iperf} tool and uses the UDP protocol. Two different file sizes are considered: 16 MB and 64 MB, and a 10 MB flow of background traffic is used for the tests.

\begin{figure}[t!!!!!!!]
\centering
\epsfig{file=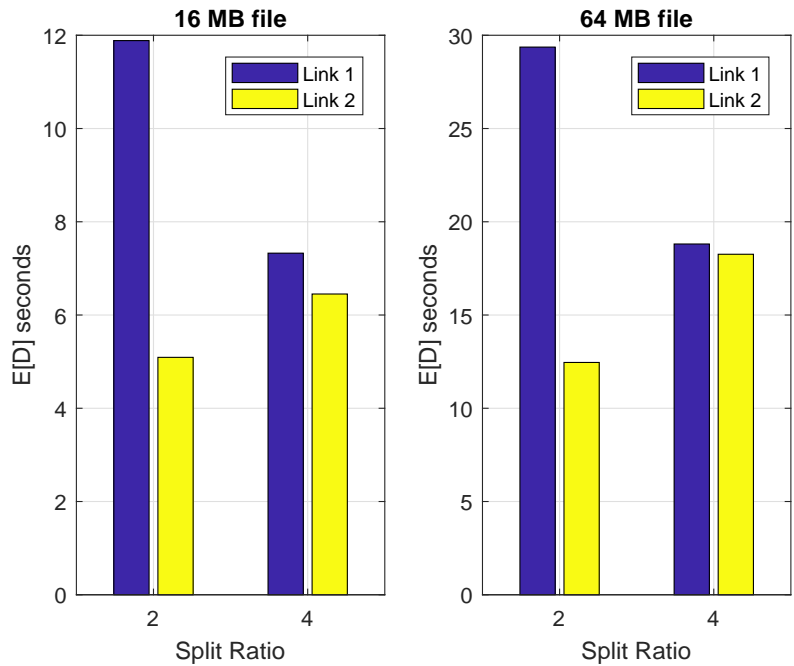,width=1\columnwidth,angle=-0}
\caption{File Transfer Time for different file sizes and background traffic loads (Third Test)}
\label{Fig:Test3}
\end{figure}

Figure \ref{Fig:Test3} shows the results obtained in this third experiment. The background traffic is injected in Link 1, reducing the available bandwidth for the file transmission. Similarly to the previous experiment, we can see that the file transfer time is improved using a fairer split ratio that takes into account the amount of background traffic. This time, we reschedule the file packets using a splitting factor of $n=4$, thus, sending the 25\% of the file only through the now busy Link 1, while the remaining 75\% is sent through the idle link. The total transfer time for 16 MB file is reduced by 30\%, and 36\% in the 64 MB file case. We also see in both cases that Link 2 transfer time is slightly increased.


\section{Performance Analysis with Multiple stations} \label{Sec:Analysis}

Once we have experimentally validated that the simultaneous use of two 802.11 interfaces associated to two different APs results in significant performance gains in terms of file transfer time when we have a single user in the network, we study if such gains can be extended when there are multiple stations sharing both APs.

\begin{figure}[t!!!!!!!!]
\centering
\epsfig{file=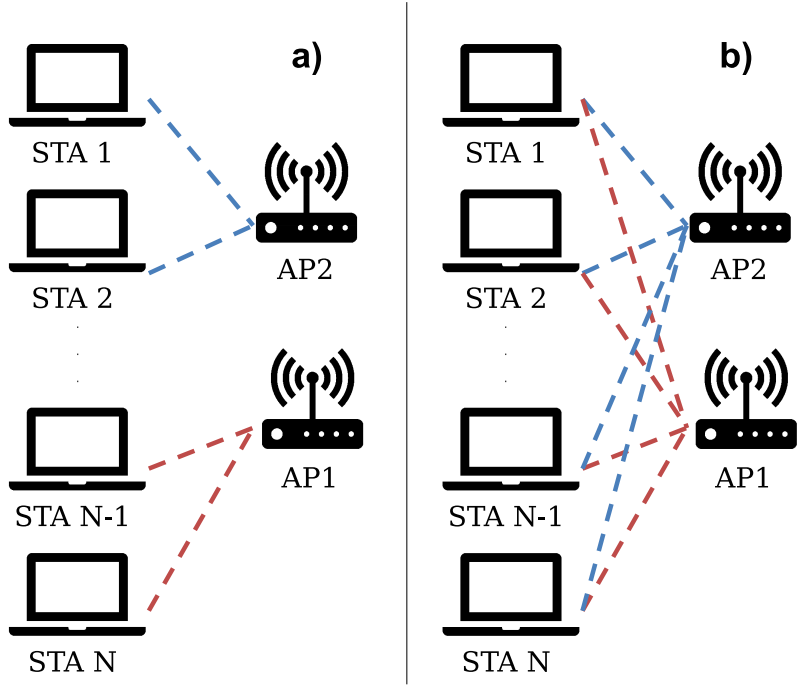,width=0.99\columnwidth,angle=-0}
\caption{The two scenarios considered in the analysis}
\label{Fig:ScenariosAnalysis}
\end{figure}

Let us consider the two scenarios depicted in Figure \ref{Fig:ScenariosAnalysis}. In scenario a), stations are equipped with a single interface, and are fairly distributed between the two APs. In case of an even number of stations, the AP to which the last station is associated is randomly selected. In scenario b), all stations are equipped with two interfaces, and are simultaneously associated to both APs. We assume that all interfaces (at the APs and stations) use always a 64-QAM modulation and a 3/4 coding rate, which for packets of size $L=12000$ results in a transmission duration of $0.253$ ms. Since the basic access scheme is employed, the duration of a collision is the same as the duration of a successful transmission. 

We consider that the two APs operate in different channels (i.e., 1 and 11, as in previous section), and all interfaces associated to a given AP are able to listen the transmissions from the others interfaces associated to the same AP. A station receives new files of average size $F$ bits to upload with rate $\lambda$ when it is idle. Then, it becomes active and starts transmitting the new file until it is completed. The file transmission rate, $\mu$, depends on the instantaneous throughput provided by the network, which depends on both the number of interfaces available at the station and the number of active stations in the network. 


\begin{figure}[h]
\centering
\epsfig{file=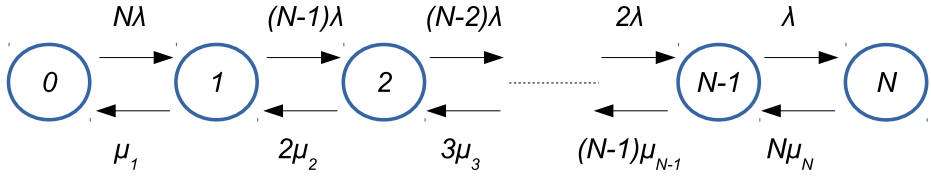,width=\columnwidth,angle=-0}
\caption{Markov chain used to characterize the number of active stations}
\label{Fig:MarkovChain}
\end{figure}

By assuming both inter-file arrivals and file transmission times are exponentially distributed, the described system can be modelled using the Markov chain shown in Figure \ref{Fig:MarkovChain}, where states represent the number of active stations, forward transition rates represent the rate at which stations become active, and backward transition rates represent the rate at which files are transmitted. The equilibrium distribution of such a Markov chain is given by
\begin{align}
	\pi_i = \frac{\prod_{j=1}^{i}{\frac{(N-j+1)\lambda}{j\mu_j(M)}}}{1+\sum_{z=1}^{N}{\prod_{j=1}^{z}{\frac{(N-j+1)\lambda}{j\mu_j(M)}}}}
\end{align}
where $M$ is the number of interfaces of each station. Backward transition rates ($\mu$) are given by $\mu_i(M)=\frac{S_i(M)}{F}$,
with 
\begin{align}
	S_i(M)= \begin{cases}
    	\frac{1}{2}\left(B_{\text{AP 1}}\left(\left \lceil \frac{i}{2} \right \rceil \right) + B_{\text{AP 2}}\left(\left \lfloor \frac{i}{2} \right \rfloor \right)  \right) & M=1 \\
    	B_{\text{AP 1}}(i) + B_{\text{AP 2}}(i) & M=2 \\        
    \end{cases},
\end{align}
the average throughput achieved by a single station when there are $i$ active  stations in the network and $M$ interfaces are used. $B_{\text{AP }n}(u)$ is the saturation throughput of a single station associated to AP $n$ computed using Bianchi's 802.11 throughput model \cite{bianchi2000performance}.

\begin{figure}[t!!!!!!!!]
\centering
\epsfig{file=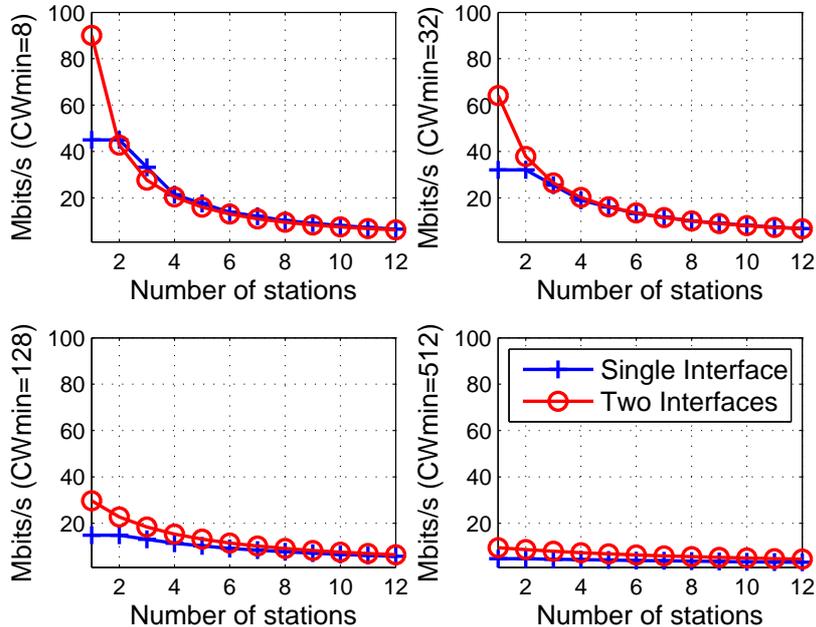,width=\columnwidth,angle=-0}
\caption{Single station throughput when the number of active stations in a WLAN increases}
\label{Fig:Figura1}
\end{figure}

\begin{figure}[t!!!!!!!!]
\centering
\epsfig{file=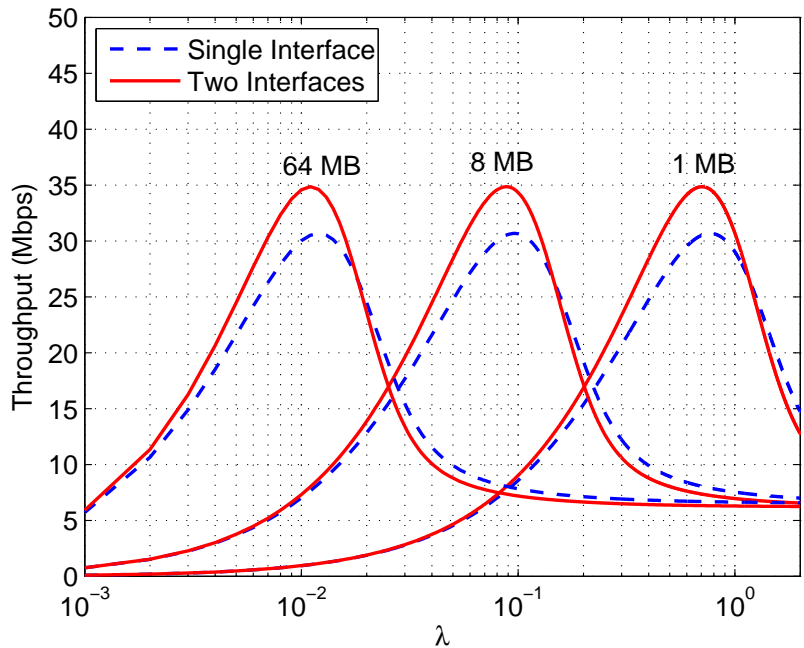,width=\columnwidth,angle=-0}
\caption{Expected per-user throughput, $E[S] = \sum_{i=0}^{N}{\pi_i S_i(M)}$}
\label{Fig:FiguraThroughput}
\end{figure}

\begin{figure}[t!!!!!!!!]
\centering
\epsfig{file=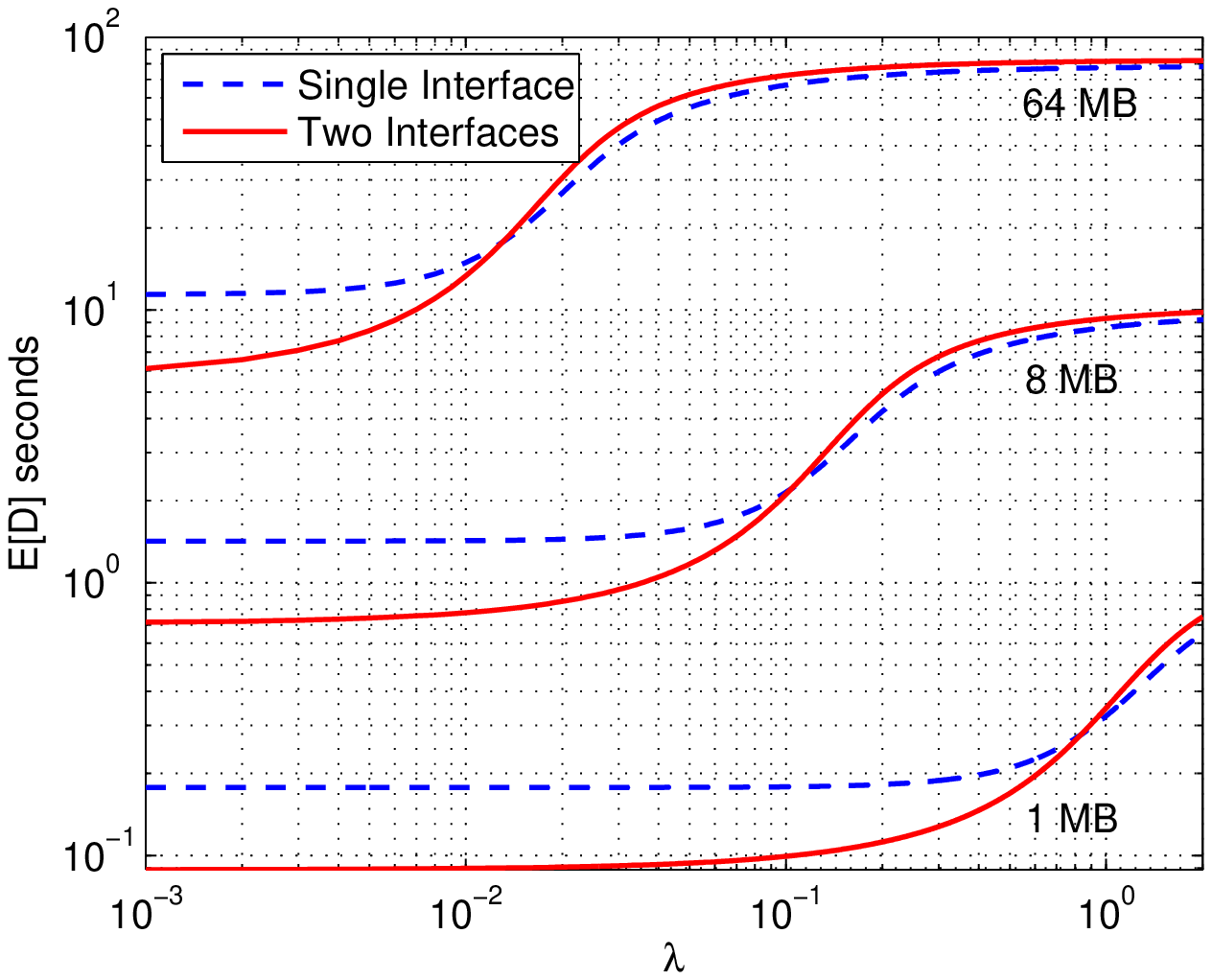,width=\columnwidth,angle=-0}
\caption{Expected file transfer time for different file sizes. The file transfer time is calculated by applying Little's formula, $E[D] = \frac{\sum_{i=0}^{N}{i\pi_i}}{\sum_{j=0}^{N}{(N-j)\lambda\pi_j}}$}
\label{Fig:FiguraDelay}
\end{figure}

Figure \ref{Fig:Figura1} shows the value of $B_{\text{AP n}}(i)$ when the number of active stations in the network increases for CW$_{\min}=8$, $32$, $128$ and $512$, and CW$_{\max}=2^5$CW$_{\min}$. We can observe that despite the relative throughput when using two interfaces is higher for larger CW$_{\min}$ values than in the case of using a single one, the highest throughput is achieved by a CW$_{\min}=8$ in both cases. For CW$_{\min}=8$, however, the use of two interfaces only results in a higher throughput when there is a single station active in the network. Otherwise, the higher collision probability when all stations use their two interfaces (i.e., it is the same as having twice the number of active stations in each AP compared to the case when a single interface is used) results in a slightly lower throughput. Therefore, the use of two interfaces will only be an interesting solution if the amount of time there is only one active station in the network is able to compensate otherwise. To dig on that situation, we plot the average per-user throughput and file transfer time when $\lambda$ increases for three different file sizes. Figures \ref{Fig:FiguraThroughput} and \ref{Fig:FiguraDelay} show that using two interfaces we are able to obtain a higher average per-user throughput and a lower file transfer time compared to the case only one interface is used for a reasonably range of $\lambda$ values. 


\section{Conclusions} \label{Sec:Conclusions}

In this paper, we have investigated if the use of multiple IEEE 802.11 interfaces is able to improve the user experience and network utilization in scenarios with multiple overlapping WLANs. To do that, we have first evaluated experimentally the time reduction when a station uploads a single file to a server using two IEEE 802.11 interfaces at the same time, compared to the case a single interface is used. We have also studied the impact of the file splitting ratio between the two interfaces, showing the importance to find a splitting ratio that balances the amount of data send through each link proportionally to its available bandwidth. 

We have complemented the experimental results by analyzing the network performance in presence of multiple contending stations using a Markovian model. The analysis done shows that the use of multiple interfaces can also be a feasible solution to improve the performance of a multi-AP network when there are several active stations, despite the higher contention that appears when multiple interfaces are used.

This paper shows promising but just preliminary and exploratory results in a topic that has to be further investigated in the next years. Next steps include testing MPTCP in high-density WLANs, including node mobility and intermittent connectivity to multiple and miscellaneous APs. Moreover, we plan to extend the analysis done to completely describe the potential gains of such an approach, and develop new protocols able to get the most of it as well.

\section*{Acknowledgment}

This  work  has  been  supported  by a Gift from the Cisco University Research Program Fund (CG\#890107, Towards Deterministic Channel Access in High-Density WLANs), a corporate advised fund of Silicon Valley Community Foundation.


\bibliographystyle{abbrv}
\bibliography{Bib}

\end{document}